\documentclass[%
 reprint,
superscriptaddress,
 amsmath,amssymb,
 aps,
]{revtex4-2}
\usepackage{graphicx}% Include figure files
\usepackage{dcolumn}% Align table columns on decimal point
\usepackage{bm}
\usepackage{color}

\newcommand{\cre}{\color{black}}

\newcommand{\rr}{{\bf r}}
\newcommand{\vv}{{\bf v}}

\newcommand{\BE}{\begin{equation}}
\newcommand{\EE}{\end{equation}}

\begin{document}

\title{Collective Vortical Motion and Vorticity Reversals \\
of Self-Propelled Particles on Circularly Patterned Substrates}
\thanks{ Physical Review E (in press)}
\author{Haosheng Wen}
\affiliation{Department of Physics and Materials Science, The University of Memphis, Memphis, TN 38152, USA}
\affiliation{Biophysics Graduate Program, The Ohio State University, Columbus, OH 43210, USA}
\author{Yu Zhu}
\affiliation{Department of Physics and Materials Science, The University of Memphis, Memphis, TN 38152, USA}
\author{Chenhui Peng}
\affiliation{Department of Physics and Materials Science, The University of Memphis, Memphis, TN 38152, USA}
\affiliation{Department of Physics, University of Science and Technology of China, Hefei, Anhui 230026, China}
\author{P.B. Sunil Kumar}
\affiliation{Department of Physics, Indian Institute of Technology Palakkad, Palakkad 668557, Kerala, India}
\affiliation{Department of Physics, Indian Institute of Technology Madras, Chennai 600036, Tamil Nadu, India}
\author{Mohamed Laradji}
\email{Corresponding author.
mlaradji@memphis.edu}
\affiliation{Department of Physics and Materials Science, The University of Memphis, Memphis, TN 38152, USA}

\begin{abstract}
The collective behavior of self-propelled particles (SPPs)  under the combined
effects of 
a circularly patterned substrate and circular confinement is investigated  through coarse-grained molecular dynamics simulations of polarized and disjoint ring polymers. {   The study is performed over a wide range of values of the SPPs packing fraction $\bar\phi$, motility force $F_D$, and area fraction of the patterned region.}
At low packing fractions, the SPPs are excluded from the system's center and exhibit a vortical motion that is dominated by the substrate at intermediate  {   values of $F_D$}. {   This exclusion zone is due to the coupling between the driving force and torque induced by the substrate, which induces an outward spiral motion of the SPPs}. {   For high values of $F_D$, the SPPs exclusion from the center is dominated by the confining boundary. At high  {   values of} ${\bar\phi}$, the substrate pattern leads to reversals in the vorticity, which become quasi-periodic with increasing ${\bar\phi}$}. {   We also found that the substrate pattern is able to separate SPPs based on their motilities.}

\end{abstract}

\maketitle

\noindent

\section{Introduction}
Active matter systems, which are collections of individual self-driven units that consume energy from the environment to move, 
have been the subject of a significant amount of research over the last few decades~\cite{Sriram2010, Sriram2017,Bechinger2016}.
Active matter systems range widely from macroscopic systems, including  schools of fish~\cite{scott2001}, flocks of birds~\cite{hemelrijk12}, and  granular media~\cite{blair2003vortices,kudrolli2008swarming,lam2015self}, to microscopic systems including colonies of bacteria, eukaryotic cells~\cite{vicsek1995novel, szabo2006phase, wang2011spontaneous, mehes2014collective}, 
actin filaments and microtubules that are propelled by their respective
motor proteins~\cite{needleman17}, and active colloids~\cite{theurkauff12}. Active matter systems often exhibit intriguing collective behavior characterized by clustering of the units and large-scale collective motion~\cite{schweitzer07}.  This collective behavior is used, for example, in bacteria colonies to reduce competition for nutrients, accelerate growth of the colony, or to increase resilience in hostile environments~\cite{kaiser07}. Likewise,  the collective behavior of assemblies of eukaryotic cells, such as epithelial monolayers and cancer cells, has physiological and pathological implications. These include embryogenesis, wound healing and tumor metastasis~\cite{gov2007collective, li2016modeling,lintz2017mechanics}.

 Many studies have shown that clustering and collective motion of self-propelled particles (SPPs) are influenced by various physical factors, including the packing fraction of the SPPs, nature of the coupling between neighboring SPPs, and the type of motion of a single SPP~\cite{Marchetti2013,Cates2015}. Other physical  factors 
include environmental constraints~\cite{Bechinger2016} such as anisotropy of the embedding fluid~\cite{doi:10.1126/science.aah6936,Song2021}, geometric confinement~\cite{Wensnick2008,vedula2012emerging,wioland2013confinement,lushi14,zuiden16,elgeti15,velasco17,Wioland16,Duclos18,Kempf19,Jain20,Norton18,Gorce19} and obstacles~\cite{Kaiser2012,aronson22}.  An interesting effect of circular confinement, for example, is an induced vortical motion of the SPPs that is concentric with the boundary ~\cite{wioland2013confinement,lushi14,zuiden16}.

 While in most studies of SPPs' collective behavior, the substrate is non-patterned, the effect of patterned substrates on SPPs collective behavior has recently  been investigated in {\cre a} few studies.  For example, it was shown that patterning the substrate, into periodic linear furrows, aligns {\em Pseudomonas aeruginosa}  along the furrows while greatly supresses their migration across them~\cite{Gloag17}. Likewise, collective migration of epithelial cells is substantially promoted by linear grooves of patterned substrates~\cite{Nam16,Lee18}.
However, computational investigations of the effect of patterned substrates on SPPs collective behavior are lacking.  
In this article, we address the effect of substrates, which are partially circularly patterned,  on the collective behavior of soft 
SPPs with the ability to switch their polarity.
In particular, we investigate how a circular confinement that is concentric with the substrate's pattern further influences their collective motion.

\section{Model and Method}

We consider a total number of ${\cal P}$ SPPs in two dimensions, each modeled as a semi-flexible ring polymer composed of $N$ beads in a good solvent. This model was recently introduced by us to investigate SPPs collective behavior  on a non-patterned substrate~\cite{Wen22} and is a generalization of an earlier model for strongly adsorbed disjoint ring polymers~\cite{Zhu21}.  The  potential energy of the SPPs is given by 
\begin{eqnarray}\label{eq:net_potentia}
&{\cal U}_{net}&={   \sum_{l=1}^{\cal P}\biggl{[}  \sum_{i=1}^N{\cal U}_{bond}\left(r^{(l)}_{i,i+1}\right)
+ \sum_{i=1}^N{\cal U}_{bend}({ \bf r}^{(l)}_{i-1},{\bf r}^{(l)}_{i},{\bf r}^{(l)}_{i+1})}\nonumber \\
 &+&{  \sum_{i=1}^N {\cal U}_{wall}\left(r^{(l)}_{i}\right)+{\cal U}_{area}\left(\{\rr^{(l)}_{i}\}\right)
 +{\cal U}_{sub}\left(\rr^{(l)}_{p_1},\rr^{(l)}_{p_2}\right)\biggl{]}}\nonumber \\
 &+&{   \sum_{l_1,l_2}\sum_{i, j}{\cal U}_{rep}\left(|\rr_i^{(l_1)}-\rr_{j}^{(l_2)}|\right),} 
\end{eqnarray}
where $\rr_i^{  (l)}$ is the coordinate of bead $i$ {   belonging to SPP $l$} and {   $r^{(l)}_{i}=|{\bf r}^{(l)}_i |$}. The $l^{th}$ SPP has two {   symmetrically positioned} poles with indices {  $p_1=1$ and $p_2=N/2+1$}.
${\cal U}_{bond}$ is a harmonic potential ensuring connectivity between consecutive beads within an SPP and is given by
\BE
{   {\cal U}_{bond}\left(r^{(l)}_{i,i+1}\right)=\frac{1}{2}k\left(r^{(l)}_{i,i+1}-r_{b}\right)^2, }
\label{eq:bond-potention}
\EE
where $k$ is the spring constant, $r^{(l)}_{i,i+1} = |\rr^{(l)}_{i+1}-\rr^{(l)}_i|$  and $r_b$ is the preferred bond length. {   In Eq.~(\ref{eq:bond-potention}), $r^{(l)}_{N,N+1}=r^{(l)}_{N,1}$.}
The semi-flexibility of an SPP's boundary is maintained through a three-body interaction 
\BE
{  {\cal U}_{bend}({ \bf r}^{(l)}_{i-1},{\bf r}^{(l)}_{i},{\bf r}^{(l)}_{i+1})= \kappa \left(1-\cos\theta^{(l)}_{i}\right), }
\label{eq:bend-potential}
\EE
where $\kappa$ is the bending stiffness of the polymers and {  $\cos\theta^{(l)}_{i} = \rr_{i-1,i }^{(l)}\cdot\rr_{i+1,i}^{(l)}/r_{i-1,i }^{(l)}r_{i+1,i}^{(l)}$}. 
{   In Eq.~(\ref{eq:bend-potential}), ${\bf r}^{(l)}_0={\bf r}^{(l)}_N$ and ${\bf r}^{(l)}_{N+1}={\bf r}^{(l)}_1$.}
Eq.~(\ref{eq:bend-potential}) 
implies that the preferred bending angle of a triplet is $180^\circ$. To account for the polarization of the SPPs, triplets of beads centered at the pole beads with indices $p_1$ and $p_2$ have a preferred bending angle $\theta_p\leq 180^\circ$. 
Since Eq.~(\ref{eq:bend-potential}) does not allow for preferred angles different from $180^\circ$,  beads $p_1$ and $p_2$ are assigned the following slightly different three-body interaction, which  allows for any arbitrary splay angle $\theta_s$,  
\BE
{   {\cal U}_{bend}({ \bf r}^{(l)}_{p-1},{\bf r}^{(l)}_{p},{\bf r}^{(l)}_{p+1})= \frac{1}{2} \kappa'\left(\cos\theta^{(l)}_{p}-\cos\theta_{s}\right)^2,}
\label{eq:bend-potential-poles}
\EE
where $\kappa'$ is the bending stiffness at the poles. Due to the softness of the potential given by Eq.~(\ref{eq:bend-potential-poles}), we found that achieving the same persistence length of the polymer with this potential as with that given by Eq.~(\ref{eq:bend-potential}) requires $\kappa'\approx 10 \kappa$.

The disjointness of the ring polymers is maintained by the following fully repulsive two-body interaction between any two non-bonded beads 
\begin{equation}
	{\cal U}_{rep}\left(r\right)=
	\left\{\begin{matrix}
		\frac{1}{2}\zeta\left(1-\frac{r}{r_{c}}\right)^2  & \text{if} & r\leq r_{c}, \\
		0 & \text{if} & r>r_{c},
	\end{matrix}\right.
	\label{eq:two-body}
\end{equation}
where $\zeta$  and $r_c$ are the strength and  range of the repulsive interaction, respectively.  Finally, the area constraint of each SPP is maintained by the effective potential energy 
\BE 
{  {\cal U}_{area}\left(\{\rr^{(l)}_{i}\}\right)=\frac{1}{2}\chi A_0\left(1-\frac{A\left(\{\rr^{(l)}_{i}\}\right)}{A_0}\right)^2, }
\label{eq:area}
\EE
where $\chi$ is the area-stretch modulus, $A_0$ is the SPP's preferred area, and {   $A\left(\{\rr^{(l)}_{i}\}\right)$} is the area enclosed by the SPP's boundary and depends on the coordinates of the beads belonging to the SPP through the shoelace formula,
\BE
{   A\left(\{\rr^{(l)}_{i}\}\right)=\frac{1}{2}\biggl{|} \sum_{i=1}^{N}\left({x^{(l)}_i}{y^{(l)}_{i+1}}-{x^{(l)}_{i+1}}y^{(l)}_{i}\right)\biggl{|},}
\label{eq:shoe-lace}
\EE
{   with $x^{(l)}_{N+1}=x^{(l)}_1$ and $y^{(l)}_{N+1}=y^{(l)}_i$.}

Finally,  the SPPs are confined within a circle of radius $R$ by the interaction potential
\begin{equation}
	{\cal U}_{wall}\left(r\right)=
	\left\{\begin{matrix}
		\varepsilon_{wall}\left(r-R+a\right)^n/a^n  & \text{if} & R-a\leq r< R, \\
		0 & \text{if} & r<R-a,
	\end{matrix}\right.
	\label{eq:cell-wall interaction}
\end{equation}
where $\varepsilon_{wall}$ and $a$ are the strength and range of this interaction, respectively. We choose $n=4$ since this value is large enough to prevent the SPPs from crossing the circular confining wall. 
The main difference between this model and prior models for the collective behavior of elongated self-propelled particles is that the present model accounts for the elongation of the self-propelled particles and their flexibility. This is in contrast with previous studies wherein particles are either rigid~\cite{peruani06,yang10,theers18} or deformable with high aspect ratio and with practically no account for the enclosed volume of the particles~\cite{duman18}.

\begin{figure}[t]
  \begin{center}
	\includegraphics[scale=0.4]{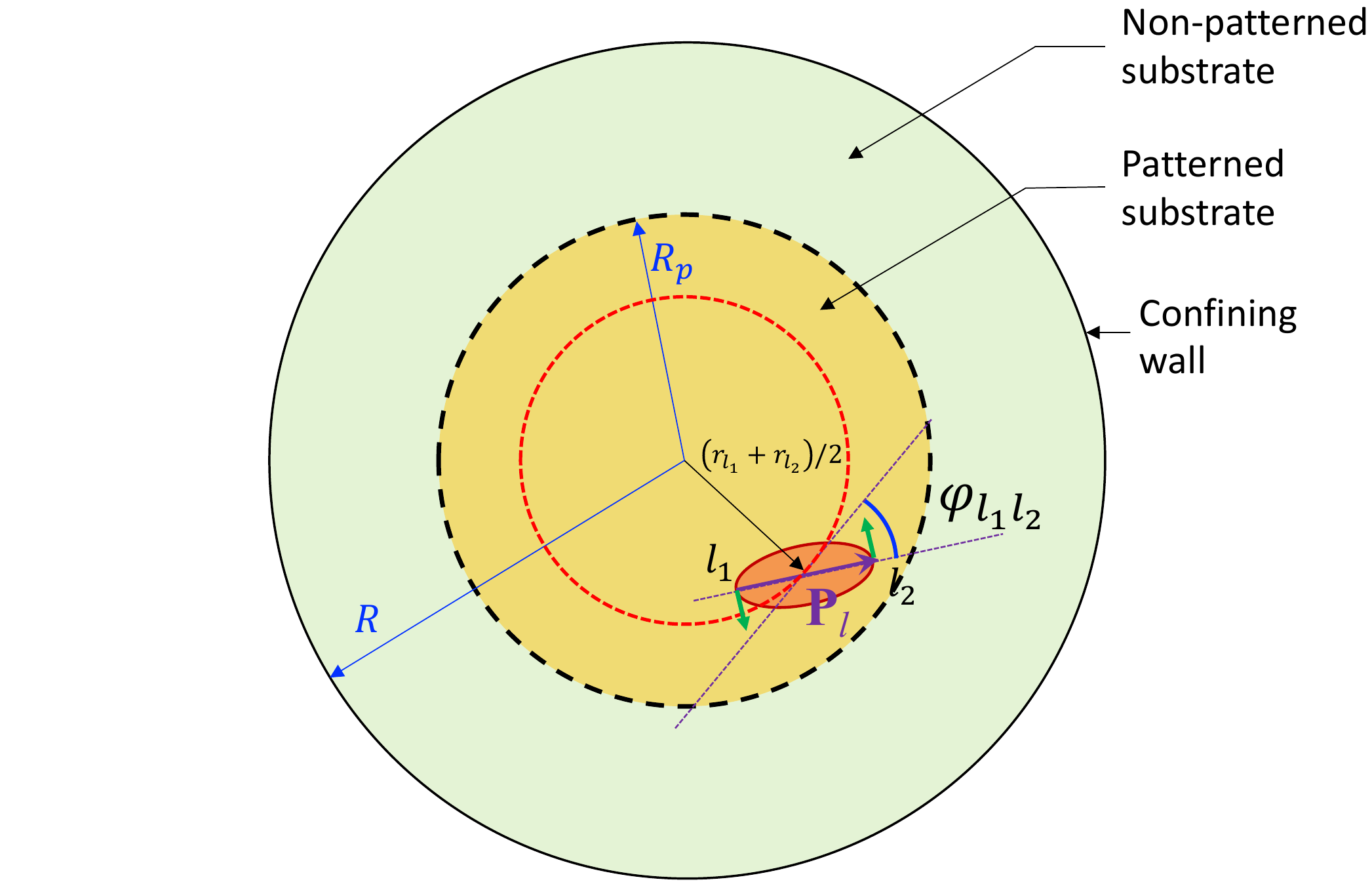}
   \end{center}
\caption{Schematic of the system. Solid black circle of radius $R$ corresponds the confining wall of the system. The yellow disk of radius $R_p$ corresponds the region of the substrate that is patterned. The green annulus corresponds to the region of the substrate that is non-patterned.
The mid-point of the polarity vector ${\bf P}_l$ of an arbitrary SPP $l$ is at a distance $(r_{l_1}+r_{l_2})/2$, where ${\bf r}_{l_1}$ and ${\bf r}_{l_2}$ are the coordinates of the two poles (The origin of the coordinate system is at the center of the system). The effect of the patterned substrate is to reorient the SPP's polarity through a torque, whose forces are indicated by the green vectors).  In this schematic, the size of the SPP is not to scale with the system size and the size of the patterned region.}
\label{fig:schematic}
\end{figure}

We consider the case where a region of the substrate is circularly patterned. Experimentally, this would correspond, for example, to a substrate that is circularly grooved~\cite{Nam16,Lee18}.  The effect of the substrate's pattern on an SPP is to align it along the local direction of the pattern. This is achieved by a simple effective potential energy between the SPP's poles that produces a torque on the SPP,
\begin{equation}
{   {\cal U}_{sub}\left(\rr^{(l)}_{p_1},\rr^{(l)}_{p_2}\right)= \frac{k_{s}}{2}\sin^2\varphi_{l},}
\label{eq: substrate interaction}
\end{equation}
where $k_{s}$ is the strength of the interaction and $\varphi_{l}$ is the angle between the polarity  ${\bf P}_l=\rr^{(l)}_{p_2}-\rr^{(l)}_{p_1}$ and the local tangent to a circle of radius $(r^{(l)}_{p_1}+r^{(l)}_{p_2})/2$ centered at the origin.  This torque 
tends to align an SPP's polarity with the local tangent of a circle centered at the origin and passing by the mid-point of the two poles, as schematically shown by Fig.~\ref{fig:schematic}.
We focus on the case where the substrate is patterned only within the region ($r\leq R_p$). Otherwise, the substrate is uniform (non-patterned) for $R_p< r\leq R$.

Each SPP is propelled by a  motility force of magnitude $F_D$, along its polarity,  that is given by
\begin{equation}
{\bf f}_l(t) = F_D \left({{\bf P}_l(t)}/{P_l(t)}\right) g\left(\bar{\bf v}_l(t),{\bf P}_l(t)\right),
\label{eq:motility_force}
\end{equation}
where $g({\bf A},{\bf B})$ = +1 or -1 if ${\bf A}\cdot{\bf B} >0$ or $<0$, respectively, and where $\bar{\bf v}_l(t)$ is the SPP's average velocity  over the time interval $\left[t-\tau_m,t\right]$, i.e. 
\begin{equation}\label{eq:SPP_velocity}
\bar{\bf v}_l(t)= \frac{1}{\tau_{m}}\int_{t-\tau_{m}}^t{\bf v}_l(t')dt',
\end{equation}
with ${\bf v}_l(t)=(1/N)\sum_{i=1}^N {\bf v}^{(l)}_i(t)$. In Eq.~(\ref{eq:SPP_velocity}), we take $\tau_m=\tau$ where $\tau=r_b\sqrt{\mu/\varepsilon}$, $r_b$ is the preferred bond length, $\varepsilon$ is the energy scale and $\mu$ is the bead's mass. 

Beads are moved according to a molecular dynamics scheme,
\begin{eqnarray}\label{eq:equation_of_motion}
	 \dot\rr^{(l)}_{i}(t) &=&  \vv^{(l)}_{i}(t), {\rm \ and} \nonumber \\
	{ \mu}\dot\vv^{(l)}_{i}(t) &=& -\nabla_{i}^{(l)}{\cal U}_{net}+\frac{{\bf f}_l(t)}{N} -\Gamma\vv^{(l)}_{i}(t)\nonumber \\
	& &\, \, \, \, +\Gamma\sqrt{2D}{\bf \Xi}^{(l)}_{i}(t),
\end{eqnarray}
where $\nabla^{(l)}_i=(\partial_{x^{(l)}_i},\partial_{y^{(l)}_i},\partial_{z^{(l)}_i})$ and ${\bf v}^{(l)}_i$ is the instantaneous velocity of bead $i$ belonging to SPP $l$. In Eq.~(\ref{eq:equation_of_motion}),  $\Gamma$ is the friction coefficient, $D$ is the diffusion coefficient of the beads in the ideal limit (i.e. in the absence of interactions and beads connectivity), and ${\bf \Xi}^{(l)}_i(t)$ is a random vector that has zero-mean and is $\delta$-correlated for the same particle and same component, i.e. ${\bf \Xi}^{(l)}_i(t)$ satisfies
\begin{eqnarray}
	\langle{\bf\Xi}^{(l)}_{i}(t)\rangle  &=&  0,\nonumber \\
	\langle{\Xi}^{(l_1)}_{i,\alpha}\left(t\right){\Xi}^{(l_2)}_{j,\beta}\left(t'\right)\rangle  &=& \delta_{l_1l_2}\delta_{ij}\delta_{\alpha\beta} \delta\left(t-t'\right),
\end{eqnarray}
where  $\alpha,\ \beta=x$ or $y$,   $\delta_{nm}$ is the Kronecker delta, and $\delta(t)$ is the Dirac delta-function.

\begin{figure*}[t]
  \begin{center}
	\includegraphics[scale=0.68]{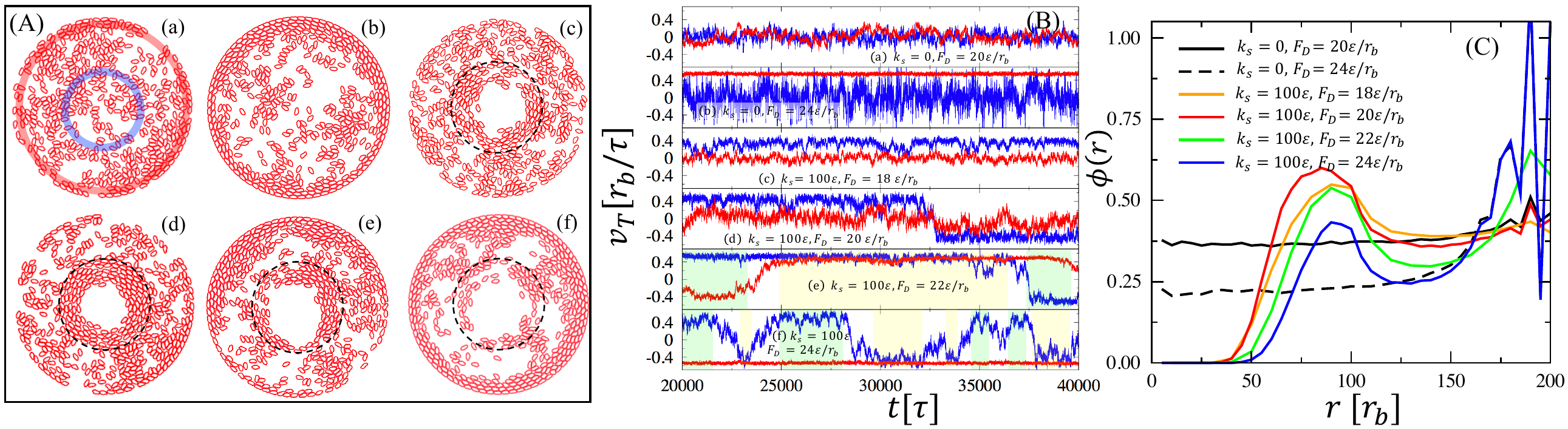}
   \end{center}
\caption{
Panel (A): Steady-state snapshots at (a) $F_{D}=20\varepsilon/r_{b}$ and $k_s=0$, (b) $F_D=24\varepsilon/r_{b}$ and $k_s=0$, (c) $F_D=18\varepsilon/r_{b}$ and $k_s=100\varepsilon$, (d) $F_D=20\varepsilon/r_{b}$ and $k_s=100\varepsilon$, (e) $F_D=22\varepsilon/r_{b}$ and $k_s=100\varepsilon$, and (f)  $F_D=24\varepsilon/r_{b}$ and $k_s=100\varepsilon$. Panel B: Time dependence of the average tangential velocity for different values of $k_s$ and $F_D$ corresponding to those in Panel (A). The blue (red) graphs correspond to SPPs in the blue (red) annulus, shown in snapshot (A). Shaded yellow (green) region corresponds to the regime where the vortices in the patterned and non-patterned regions are in same (opposite) directions. Panel (C):
Radial profiles of the packing fraction, ${\bar\phi}$  at values of $F_D$ and $k_s$ corresponding to those in Panel (A). All data shown in this figure are at  ${\bar\phi}=0.398$, $R_p=100r_b$ and $R=200r_b$.}
\label{fig:snapshots_density}
\end{figure*}
The equations of motion are integrated using the velocity-Verlet algorithm with a time step $\Delta t=0.01\tau$.  The numerical value of a component of the random force is  given by
\BE
\Xi_{i,\alpha}^{(l)}=\left(\frac{3}{\Delta t}\right)^{1/2} \lambda_{i,\alpha}^{(l)},
\EE
where $\lambda_{i,\alpha}^{(l)}$ is a random number generated from a uniform distribution in the interval $[-1,1]$.
Each SPP is composed of $N=40$ beads. The values of the parameters of the model SPPs, {   which are kept fixed}  in the present study, are given by
\begin{eqnarray}\label{eq:parameters}
         k  =  100\varepsilon/r_b^2,\,  \kappa  &=&  100 \varepsilon,\,   \kappa'  =  1000 \varepsilon,\, \theta_s=120^{\circ},\,  \zeta=50\varepsilon, \nonumber\\
         r_c&=&r_b,\,  \chi  =  1\varepsilon/r_b^2,\,  A_0  =  100 r_b^2,\, \tau_m=\tau,\,  \nonumber \\
         D&=&1.0r_b^2/\tau,\, {\rm and}\, \,  \Gamma=1.0 { \mu}/\tau  .
\end{eqnarray}

\section{Results}

\subsection{Effects of Patterned Substrate and Motility Force on SPPs' Collective Behavior}
We first focus on the combined effect of the patterned substrate and circular confining wall on the SPPs collective behavior at an average 
packing fraction ${\bar\phi}={\cal P}A_0/\pi R^2=0.398$ with $R=200 r_b$. This corresponds to ${\cal P}=500$.
Steady-state snapshot (a) in Fig.~\ref{fig:snapshots_density}(A) and Movie 1, 
at $F_D=20\varepsilon/r_b$ and non-patterned 
substrate ($k_s=0$), indicate a small amount of clustering and a weak collective motion, in agreement with  prior results~\cite{Wen22}. Fig.~\ref{fig:snapshots_density}(C) shows that at these conditions, the radial distribution of the SPPs packing fraction, $\phi(r)$, is almost uniform. As $F_D$ is increased to $24\varepsilon/r_b$ at $k_s=0$, the motility force drives many SPPs to the boundary leading to their accumulation as shown by snapshot (b) in Fig.~\ref{fig:snapshots_density}(A) and collective unidirectional vortical motion (see Movie 2). This is also demonstrated by the time dependence of the average tangential velocity of the SPPs in an annulus of thickness $10r_b$ near the boundary (red graph in Fig.~\ref{fig:snapshots_density}(B) at $k_s=0$ and $F_D=24\varepsilon/r_b$). In contrast, the SPPs motion in an annulus close to the center is fairly turbulent (blue graph in Fig.~\ref{fig:snapshots_density}(B) at $k_s=0$ and $F_D=24\varepsilon/r_b$). SPPs accumulation at the boundary is due to the asymmetry between the effect of the motility force, which drives the SPPs toward the boundary, and thermal effects, which drive the SPPs away from the boundary, and has been observed in earlier studies~\cite{Bechinger2016}. 
In contrast, although the SPPs that are away from the boundary move collectively in clusters, they do not exhibit a net vortical motion, as demonstrated by the fluctuations around 0 of the average tangential velocity of the SPPs in the annulus close to the center (blue graph in Fig.~\ref{fig:snapshots_density}(B) at $k_s=0$ and $F_D=24\varepsilon/r_b$). 

Interaction between the SPPs and the patterned substrate leads to a much richer dynamical behavior.  Snapshots (c) and (d) in Fig.~\ref{fig:snapshots_density}(A) and their corresponding tangential velocities vs. time in Fig.~\ref{fig:snapshots_density}(B) show that, at $k_s=100\varepsilon$ and $F_D=18$ or $20\varepsilon/r_b$, the patterned substrate and the driving force collectively lead to (1) a tangential alignment of the SPPs in the patterned region, (2) their accumulation at the periphery of the patterned region, and (3) their exclusion from the center. At $k_s=100\varepsilon$ and $F_D=20\varepsilon/r_b$,
Fig.~\ref{fig:snapshots_density}(B) and Movie 3 show that the SPPs move as a vortex, in the patterned region of the substrate, with very few reversals in its direction.
In contrast, the SPPs outside the patterned region exhibit a weak collective behavior, as demonstrated by the fact that the SPPs' average tangential velocity in this region fluctuates around 0 (red graph in Fig.~\ref{fig:snapshots_density}(B) at $k_s=100\varepsilon$ and $F_D=20\varepsilon/r_b$).

As $F_D$ is further increased to $F_D=22$ or $24\varepsilon/r_b$, at $k_s=100\varepsilon$, the corresponding snapshots (d) or (e), respectively,  shown in Fig.~\ref{fig:snapshots_density}(A), show that more SPPs are driven to the confining wall. This is also demonstrated by increased packing fraction next to the boundary at these values of $F_D$  in Fig.~\ref{fig:snapshots_density}(C). Fig.~\ref{fig:snapshots_density}(B) shows that, at these values of $F_D$,
the SPPs exhibit collective vortical motion in both patterned and non-patterned regions. These vortices can move either in the same direction (shaded yellow regions in Fig.~\ref{fig:snapshots_density}(B) and Movie 4) or opposite directions (shaded green regions and Movie 5) with frequent reversals. Inspection of the vorticity reversals indicates that they are due to collectively moving clusters in the non-patterned region,
which collide with the vortices in the patterned region or in the boundary layer.

\begin{figure}[b]
  \begin{center}
	\includegraphics[scale=0.66]{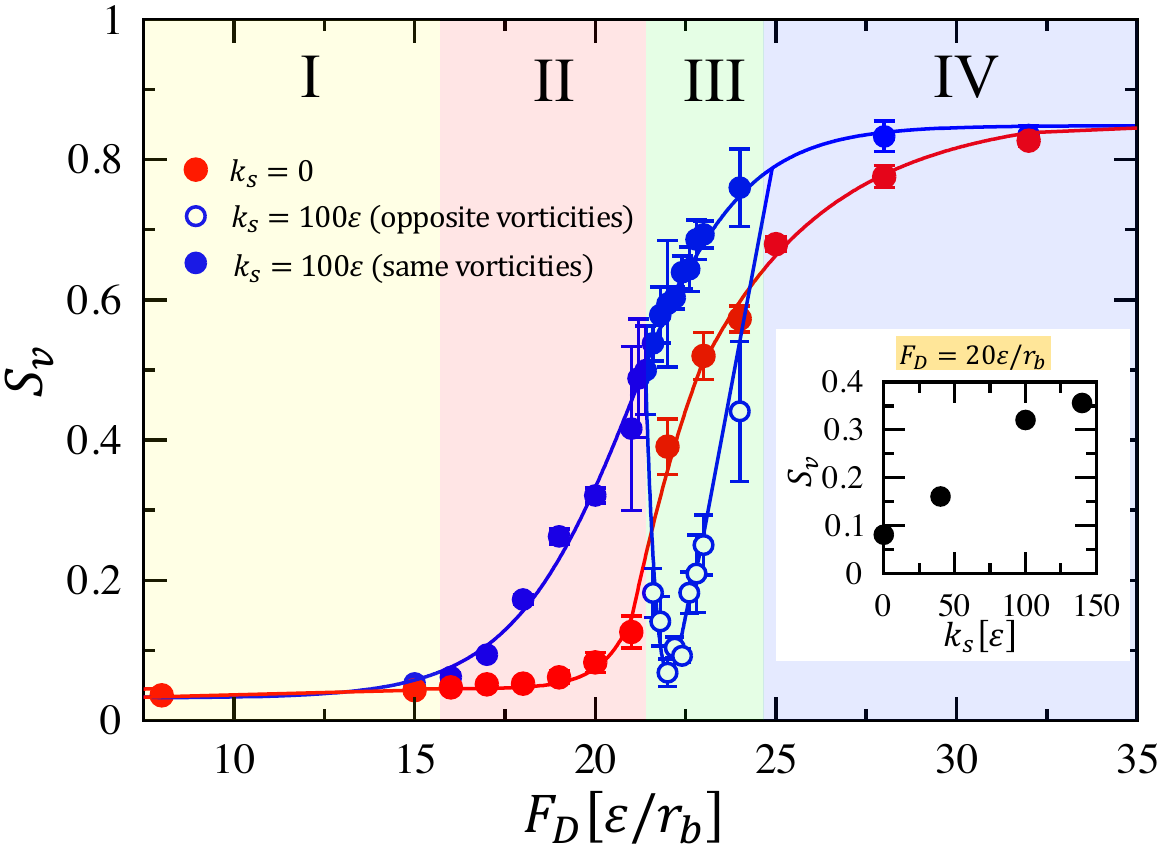}
   \end{center}
\caption{ 
$S_V$ vs. $F_D$ at ${\bar\phi}=0.398$, $R_p=100r_b$ and $R=200r_b$ for $k_s=0$ (red circles) and $k_s=100\varepsilon$ (blue circles). Full (open) blue circles correspond to $S_v$ at $k_s=100\varepsilon$ where the vortices in the patterned and non-patterned regions have same (opposite) directions.(Inset) $S_v$ vs. $k_s$ at $F_D=20\varepsilon/r_b$. The solid lines are simply guides to the eye.}
\label{fig:order-parameter}
\end{figure}

The SPPs collectivity is quantified through 
the vortical order parameter defined as
\begin{equation}
S_{v}=\langle|\sum_{l=1}^{\cal P}{\sigma_{l}}|\rangle/{\cal P}, 
\label{vorticity}
\end{equation}
where $\sigma_{l} = +1$ (-1) if the direction of the tangential velocity of SPP $l$ is clockwise (counter-clockwise). Fig.~\ref{fig:order-parameter},  which depicts $S_v$ vs. $F_D$ at ${\bar\phi}=0.398$, shows that the substrate pattern shifts the onset of vortical collective motion to smaller values of $F_D$.  Four distinct regimes in the case of $k_s=100\varepsilon$ are identified. In regime I ($F_D\lesssim 16\varepsilon/r_b$), there is no collective motion. In regime II ($16\varepsilon/r_b\lesssim F_D\lesssim 21\varepsilon/r_b$), the collective behavior is dominated by the patterned region, and is characterized by an almost unidirectional vortical motion. 
Fig.~2(C) shows that regime II is also characterized by an increase in the maximum of the SPPs packing fraction in the patterned region with increasing $F_D$. In regime III ($21\varepsilon/r_b\lesssim F_D\lesssim 25\varepsilon/r_b$), both patterned substrate and confining wall independently promote SPPs collective motion, and lead to vortical motion in both regions with same or opposite directions. This results in a bifurcation of $S_v$ into two branches: one branch with high values of $S_v$ (solid blue circles in Fig.~3) where the two vortices have same direction, and a second branch with low values of $S_v$ (open blue circles in Fig. 3) where the two vortices have opposite directions. Regime III marks the beginning of the decrease in the value of the maximum of the SPPs' packing fraction  in the patterned regions. Finally, in regime IV ($F_D\gtrsim 24\varepsilon/r_b$), the majority of the SPPs are accumulated near the confining wall, where they move as a unidirectional vortex.

\begin{figure}[t]
  \begin{center}
	\includegraphics[scale=0.66]{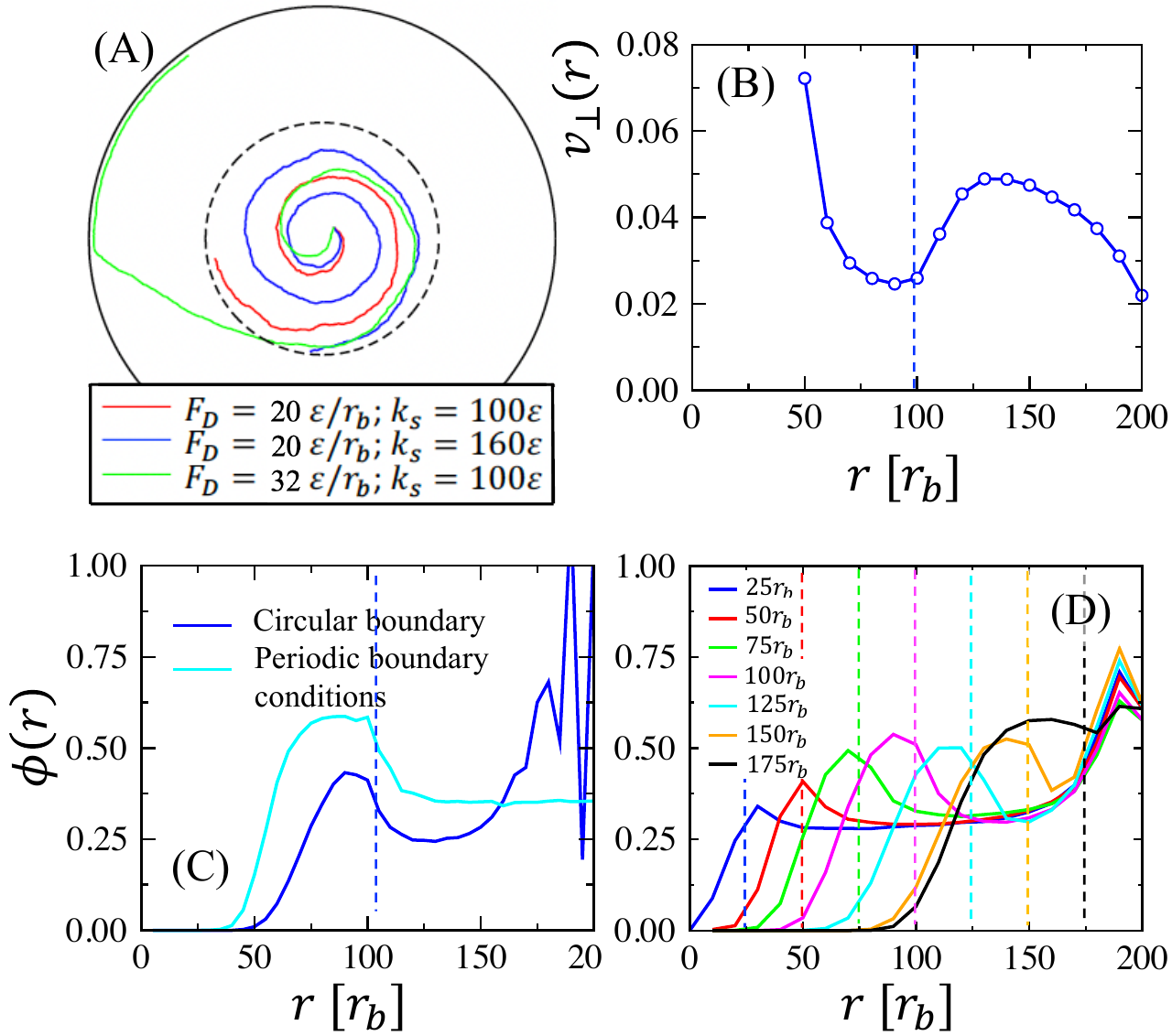}
   \end{center}
\caption{(A) trajectories of a single SPP starting from a position near the center, for differemt values of $F_D$ and $k_s$. (B) Radial profile of the radial velocity of the SPPs for the case of a circular confining wall. (B) Radial profile of the packing fraction for the case of a circular confining wall (solid line) and PBC (dashed line). Data shown in (B) and (C) are in the case of $F_D=24\varepsilon/r_b$, $k_s=100\varepsilon$, $\bar\phi=0.398$, $R_p=100r_b$ and $R=200r_b$. (D) Radial profiles of the packing fraction for different values of the radius of the patterned region, $R_p$, indicated in the legend.  These data correspond to $F_D=22\varepsilon/r_b$, $k_s=100\varepsilon$, $R=200r_b$ and $\bar\phi=0.398$. The  vertical dashed lines in (B-D) indicate the location of the boundary between the patterned (left) and non-patterned (right) regions of the substrate. }
\label{fig:spiral}
\end{figure}

Interestingly,  snapshots (c) to (f) of Fig.~\ref{fig:snapshots_density}(A) and Fig.~\ref{fig:snapshots_density}(C) demonstrate that the patterned substrate induces an exclusion zone  in the center with a diameter that increases with $F_D$. This is contrasted with the case of a non-patterned substrate, in which the radial profile of the packing fraction is almost uniform, except at the boundary. 
The source of this exclusion zone, is inferred from simulations of a single SPP (dilute regime) at finite values of $k_s$ and $F_D$, starting from a location near the center. Fig.~\ref{fig:spiral}(A) (see Movie 6 as well) shows that the SPP's trajectory is an outward spiral, with a number of turns that increases with increasing $k_s$ or decreasing $F_D$.  Hence, the motility force and the substrate's pattern cooperatively drive the SPPs away from the patterned region with a rate that increases with 
$F_D$ and decreases with $k_s$, leading to an exclusion zone in the center.

In addition to the exclusion zone in the center,  Fig.~\ref{fig:snapshots_density}(C) shows that the radial profile of the packing fraction exhibits a broad peak within the patterned region, and close to the boundary between the patterned and non-patterned regions. The emergence of this peak is understood as follows. The motion of the SPPs within the patterned region is mainly tangential, while in the non-patterned region (but away from the confining wall), the motion is more turbulent. As a result $v_\perp^p<v_\perp^n$, where $v_\perp^p$ and $v_\perp^n$ are  the {   averages of the magnitudes of the} radial components of the SPPs  velocities in the patterned and non-patterned regions, respectively, as shown in Fig.~\ref{fig:spiral}(B). Steady state requires that the outflux of the SPPs from the patterned must be equal to the influx of the SPPs from the non-patterned regions to the patterned region, i.e. $\phi_p v_{\perp,out}^p=\phi_n v_{\perp,in}^n$, where $\phi_p$ ($\phi_n$) is the packing fractions of the SPPs in the patterned (non-patterned) region, close the boundary between the patterned and non-patterned regions. 
{   $v_{\perp,out}^p$  is the average of the radial component of the velocity of the SPPs outgoing from the patterned region at the boundary between the patterned and non-patterned regions. Likewise,   $v_{\perp,in}^n$ is the average of the radial component of the velocity of the SPPs incoming from the non-patterned region at the boundary between the patterned and non-patterned regions.}
Therefore, mass balance between the outflux and influx of the SPPs across this boundary, at steady state, imposes  $\phi_p > \phi_n$. Combined with the fact that the interplay between the motility force and the torque induced by the patterned substrate, which leads to SPPs exclusion from the center, the argument above implies that the radial packing fraction profile must exhibit a peak within the patterned region, and close to the boundary between the patterned and non-patterned regions, as shown by Fig.~\ref{fig:snapshots_density}(C).
$F_D$ enhances the SPPs outflux from the patterned region, i.e. it increases $v_\perp^p$, while it decreases the influx from the non-patterned region, due to increased accumulation of the SPPs near the confining wall. As a result,
the size of the exclusion zone increases with $F_D$ (see  Fig.~\ref{fig:snapshots_density}(C)). Elimination of SPPs accumulation at the boundary, through imposing periodic boundary conditions (PBC),  enhances SPPs influx from the non-patterned region to the patterned region. This leads to a decrease in the size of the exclusion zone, as demonstrated by Fig.~\ref{fig:spiral}(C). 

The results thus far presented correspond to the case of a radius of the patterned region of the substrate, $R_p=100r_b$. To infer the effect of the size of the patterned region, we performed a series of simulations in the case of $\bar\phi=0.398$, $F_D=22\varepsilon/r_b$, $k_s=100\varepsilon$, and $R=200r_b$. 
{Fig.~\ref{fig:spiral}(D) shows the radial profile of the packing fraction of these systems with $R_p$ varying between $25 r_b$ and $175r_b$. This figure demonstrates that the diameter of the depletion zone increases with $R_p$, which implies that the size depletion of the SPPs from the middle is also affected by the behavior of the SPPs in the non-patterned region of the substrate, in line with the arguments presented in the previous paragraph. }

\subsection{Effect of SPPs' Packing Fraction on their Collective Behavior on a Patterned Substrate}
\begin{figure}[t]
  \begin{center}
	\includegraphics[scale=0.73]{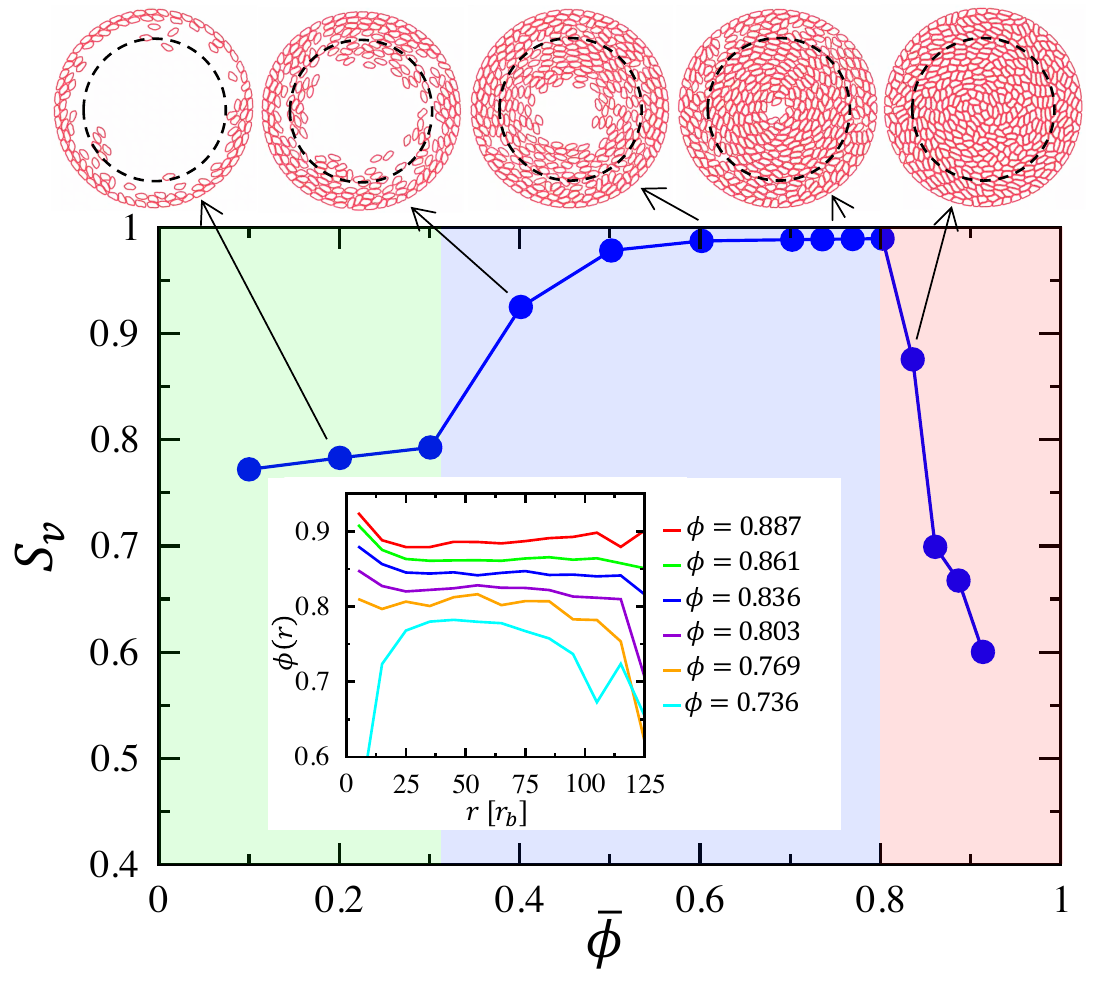}
   \end{center}
\caption{
Vortical order parameter vs. packing fraction at $k_s=100\varepsilon$, $F_D=24\varepsilon/r_b$, $R_p=100r_b$  and $R=138r_b$. Vortical motion is dominated by the circular confining wall at low ${\bar\phi}$ (green region). Both circular confining wall and 
patterned substrate contribute to vortical motion at intermediate ${\bar\phi}$ (blue region). At high ${\bar\phi}$, vortical motion exhibits reversals (red region). Inset shows radial packing fraction profiles at different values of ${\bar\phi}$. Steady state snapshots at different packing fractions are shown at the top of the figure. The dashed circles in these snapshots indicate the boundary of the patterned region of the substrate.} 
\label{fig:density_profile_order_parameter_vs_rho}
\end{figure}

We now turn to the effect of SPPs packing fraction on their collective motion. We consider the case where $F_D=24\varepsilon/r_b$ and $k_s=100\varepsilon$. The packing fraction is varied by changing the number of SPPs from ${\cal P}=59$ to 540, while the radius of the system is kept fixed at $R=138r_b$. 
Corresponding $S_v$ vs. ${\bar\phi}$, shown in Fig.~\ref{fig:density_profile_order_parameter_vs_rho}, reveals three main regimes. For 
${\bar\phi}\lesssim 0.3$, most SPPs accumulate at the boundary where they move as a unidirectional vortex 
(see Movie 7). For  $0.3\lesssim{\bar\phi}\lesssim 0.8$, the amount of SPPs is increased in the patterned region, where they move as a vortex with same direction as that in the boundary layer (see Movie 8). 
Fig.~\ref{fig:density_profile_order_parameter_vs_rho} shows  that for ${\bar\phi}\lesssim 0.8$, $S_v$ increases monotonically with ${\bar\phi}$. 
Surprisingly, however, $S_v$ decreases with ${\bar\phi}$ for ${\bar\phi}\gtrsim 0.8$. This decrease is interestingly correlated with the disappearance of the exclusion zone in the center as demonstrated by the profiles of the packing fraction in the inset of Fig.~\ref{fig:density_profile_order_parameter_vs_rho}. In fact, the inset of Fig.~\ref{fig:density_profile_order_parameter_vs_rho} shows that an excess of SPPs at the center is induced at ${\bar\phi}\gtrsim 0.8$.

\begin{figure*}[t]
  \begin{center}
	\includegraphics[scale=0.67]{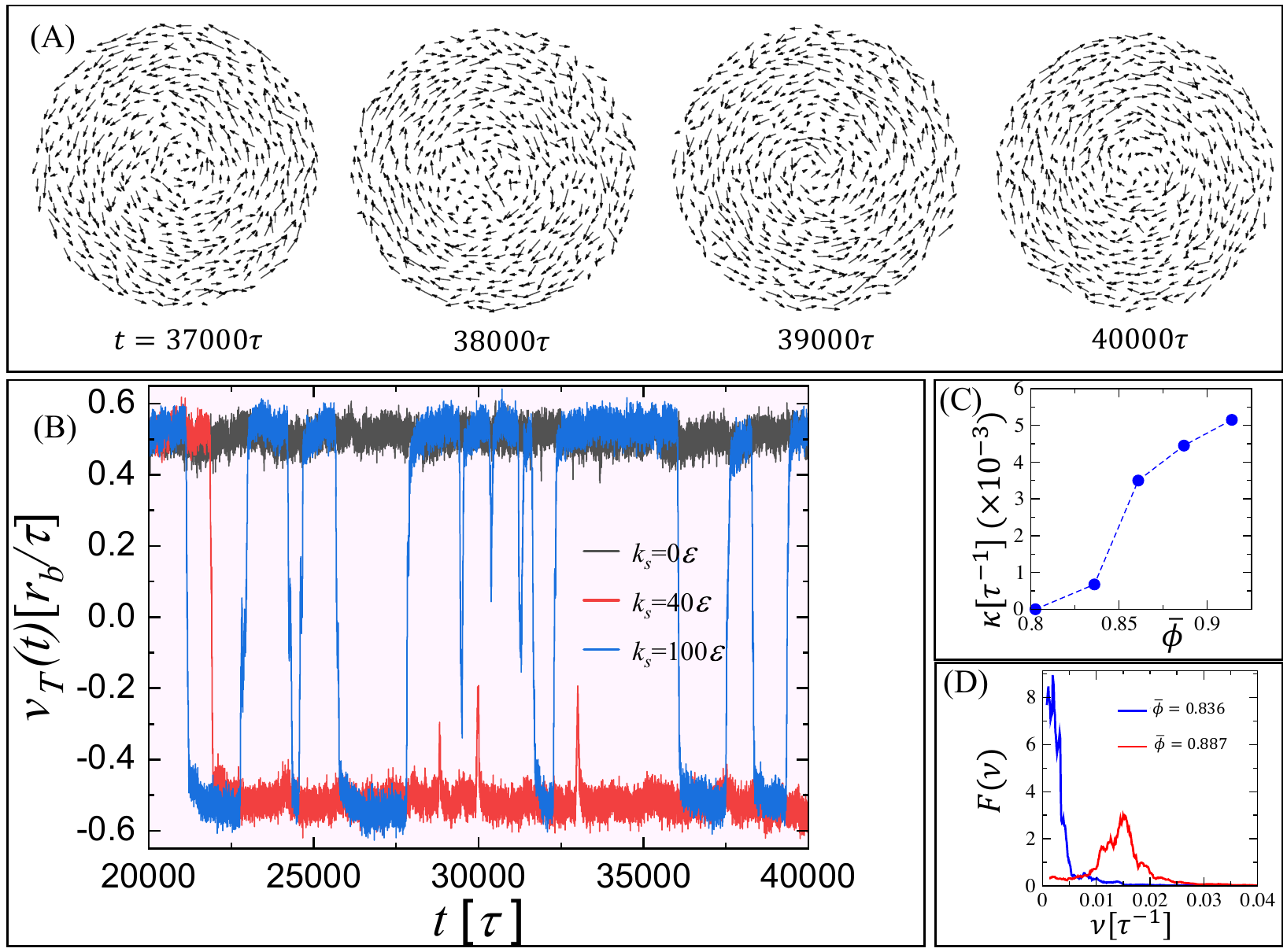}
	
   \end{center}
\caption{(A) Time-sequence of velocity  snapshots showing vorticity reversals
at $F_{D}=24\varepsilon/r_{b}$, ${\bar\phi}=0.836$, $R_p=100r_b$, $R=138r_b$ and $k_{s}=100\varepsilon$. (B) Tangential velocity ${v}_{T}(t)$ vs. time at $F_{D}=24\varepsilon/r_{b}$ and ${\bar\phi}=0.836$.  (C) Rate of vorticity reversals vs. ${\bar\phi}$ at $k_{s}=100\varepsilon$. (G) The Fourier transform, $F(\nu)$,  of the velocity autocorrelation function $f(t)=\langle v_T(t_0+t)v_T(t_0)\rangle$, vs. frequency at $k_{s}=100\varepsilon$ at two high values of the packing fraction.  }
\label{fig:reversals}
\end{figure*}

Inspection of movies at ${\bar\phi}\gtrsim 0.8$ reveals an emergence of reversals in the vorticity (demonstrated by SPPs velocities snapshots in Fig.~\ref{fig:reversals}(A) and by Movie 9). 
These reversals are quantified by the time dependence of ${ v}_{T}(t)$, defined as the average of the tangential velocity of the SPPs in an annulus of thickness $10r_b$ near the system's boundary.
Fig.~\ref{fig:reversals}(B) shows that 
${ v}_{T}$ is essentially constant in the case of  a non-patterned substrate ($k_s=0$) at $F_D=24\varepsilon/r_b$, 
indicating a unidirectional vortical motion. At $k_s=40\varepsilon$ and same $F_D$, Fig.~\ref{fig:reversals}(B) shows that ${ v}_{T}$ exhibits a single reversal during the time interval $[20\, 000\tau,\, 40\, 000\tau]$. In stark contrast, however, 
${ v}_{T}$ exhibits many reversals at  $k_s=100\varepsilon$ and same $F_D$ during the same time interval.
Therefore, at high packing fractions, the rate of vorticity reversals (i.e.,  number of reversals per unit of time), $\kappa$, increases with increasing $k_s$ beyond some threshold value. 
Likewise, Fig.~\ref{fig:reversals}(C) shows that $\kappa$ increases with ${\bar\phi}$ for ${\bar\phi}\gtrsim 0.8$. The decrease in $S_v$ at ${\bar\phi}\gtrsim 0.8$, shown in Fig.~\ref{fig:density_profile_order_parameter_vs_rho}(B), is simply due to  coexistence of two vortices with opposite directions during the reversal events, as demonstrated by a series of snapshots in Fig.~S1 in Supplemental Information~\cite{suppl}.

\begin{figure}[b]
  \begin{center}
	\includegraphics[scale=0.5]{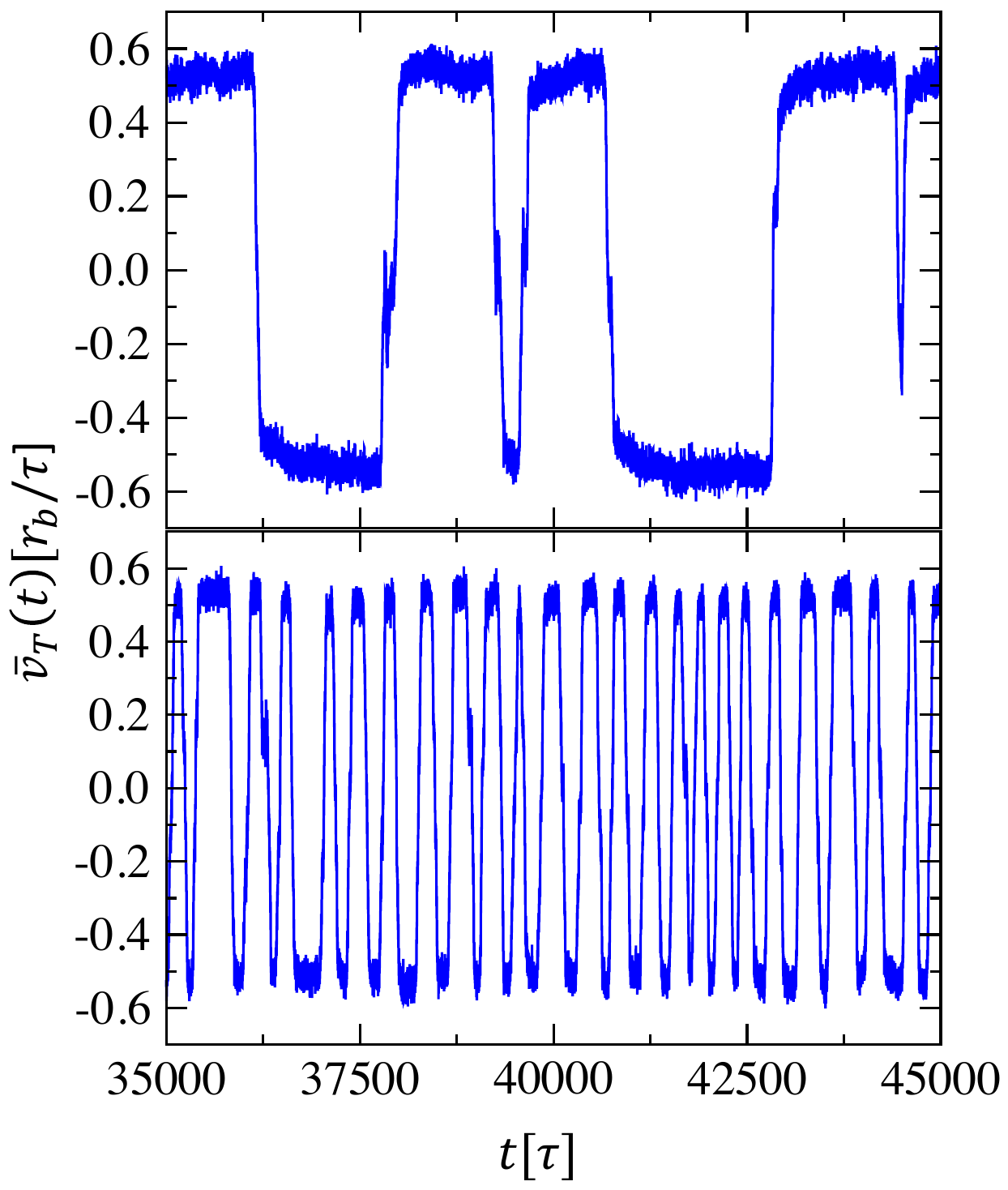}
   \end{center}
\caption{Time dependence of the tangential velocity of an annulus of thickness $10 r_b$ near the system's boundary for the case of $F_D=24\varepsilon/r_b$, $R_p=100r_b$, $R=200 r_b$ and $k_s=100\varepsilon$. Top and bottom graphs correspond to $\bar\phi=0.836$ and $0.887$, respectively.}
\label{fig:quasi-periodic}
\end{figure}

Correlations between reversal events are inferred from the power spectrum $F(\nu)$, defined as the Fourier transform of the velocity autocorrelation $f(t)=\langle v_T(t_0+t)v_T(t_0)\rangle$, where $\nu$ is frequency.
Fig.~\ref{fig:reversals}(D) shows that, at ${\bar\phi}=0.836$, $F(\nu)$ is peaked 
at $\nu\approx 0$. This  indicates that reversal events are weakly correlated at packing fractions around this value of ${\bar\phi}$. 
Fig.~\ref{fig:reversals}(D) shows that $F(\nu)$ exhibits a well-defined peak at ${\bar\phi}=0.887$. 
Therefore, reversal events of the vorticity become interestingly quasi-periodic with increasing ${\bar\phi}$. The emergence of quasi-periodic reversals at high densities is also demonstrated by the time dependence of the tangential velocity in Fig.~\ref{fig:quasi-periodic}.

Inspection of Movie 9 shows that vorticity reversals always originate from the center of the system.
This concurs with the fact that vorticity reversals are absent at low packing fractions, i.e. when  the exclusion zone is present. To demonstrate that the geometry of the confining wall has {\cre a weak effect} on vorticity reversals, we performed a simulation on a system with a square boundary, of linear size $L_x=400r_b$, and same circular pattern with $k_s=100\varepsilon$, $F_D=24\epsilon/r_b$, $\bar\phi=0.887$ and $R_p=100r_b$, and found reversals in the vorticity similar to the case with circular boundary and with about same value of $\kappa$, as demonstrated by Fig.~S2~\cite{suppl}.
Likewise, Fig.~S3~\cite{suppl} shows that systems with periodic boundary conditions, at same values of $F_D$, $k_s$, $\bar\phi$, $R_p$ and $L_x$, also exhibit vorticity reversals, albeit not as correlated as in the case of circular or square boundary. This is due to the fact the periodic boundary conditions induce more turbulent flow of the SPPs in the non-patterned region.

As stated above, reversals in the vorticity are associated with an increase in SPPs packing fraction at the center. This is found to also be associated with an increase in the misalignment between the SPPs polarities and velocities, as shown by Fig.~S4 (A)~\cite{suppl}. This results in a high degree of fluctuations in the average of the tangential  velocity of the SPPs in the center as opposed to those away from the center, as shown by Fig.~S4 (B)~\cite{suppl}. These increased fluctuations at the center leads some SPPs to move in a direction opposite to that of the vortex, and in some cases these SPPs force neighboring SPPs to follow, leading to the observed intermittent vorticity reversals. 

\begin{figure}[t]
  \begin{center}
	\includegraphics[scale=0.63]{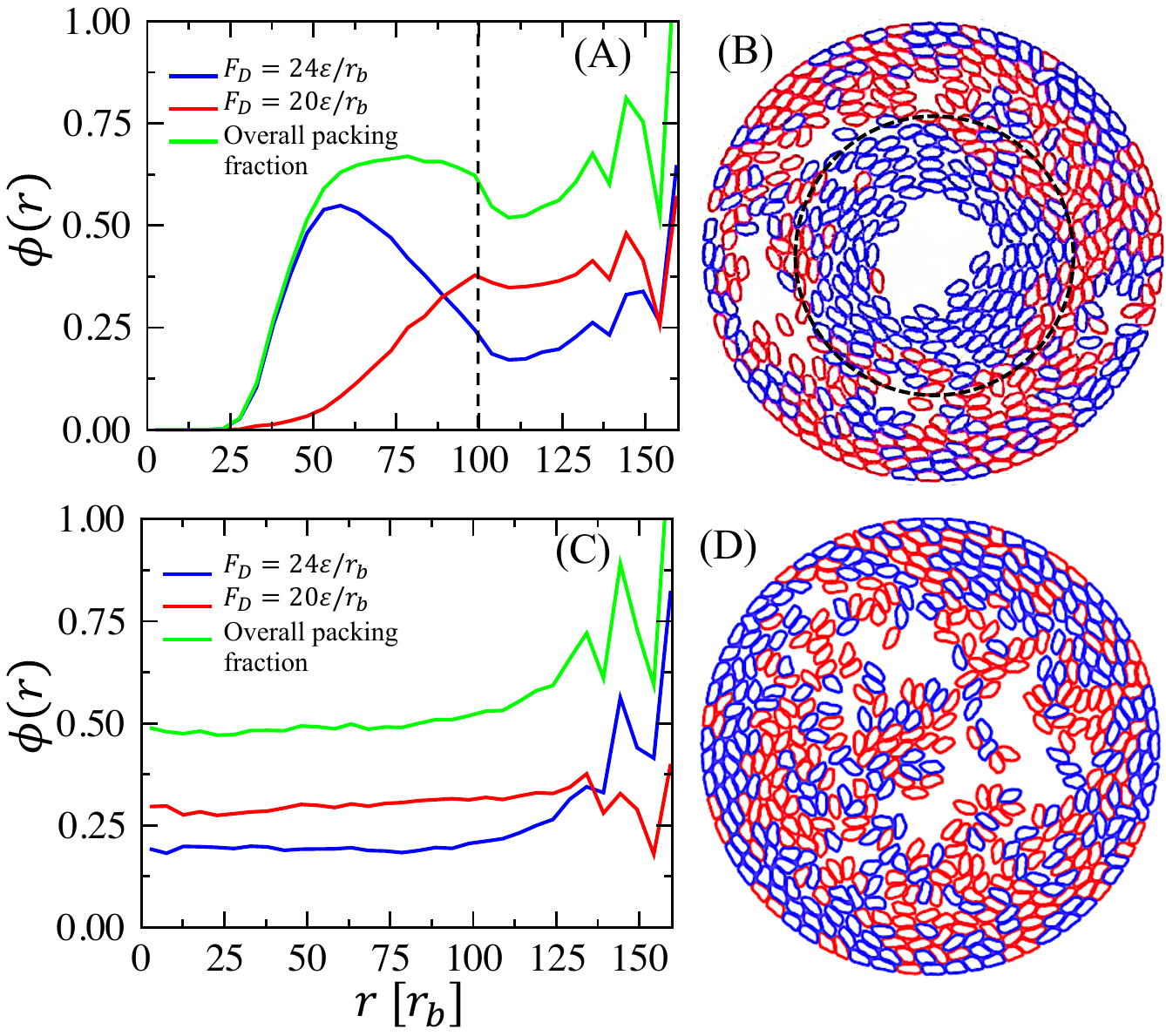}
   \end{center}
\caption{(A) Radial profile of the packing fraction in the case of a binary system of fast SPPs, with $F_D=24\varepsilon/r_b$ (blue) and slow SPPs, with $F_D=20\varepsilon/r_b$ (red), in the case where the average packing fraction is 0.6, $k_s=100\varepsilon$, $R=162r_b$ and $R_p=100r_b$. (B) A snapshot of the binary system at steady state. Blue and red SPPs correspond to fast and slow SPPs, respectively. The dashed vertical line and  circle  in (A) and (B), respectively, indicate the boundary of the patterned region.  (C) and (D) same as in (A) and (B), respectively, but in the case of a non-patterned substrate ($k_s=0$).}
\label{fig:binary}
\end{figure}

\subsection{Patterned-substrates induced segregation between fast and slow SPPs} 
Our simulations show that at low and intermediate values of the packing fraction, the SPPs spatial distribution depends on their motility force. One would therefore expect that patterning the substrate may be used as a tool to spatially separate SPPs, based on their motility force. To verify this hypothesis, we performed a simulation of a binary system,  at an average packing fraction $\bar\phi=0.6$, in which half of the SPPs are slow (with $F_d=20\varepsilon/r_b$) and the other half are fast  (with $F_D=24\varepsilon/r_b$). The two types of SPPs  are otherwise identical. The packing fraction profiles of the two components and a steady-state snapshot, depicted in Figs.~\ref{fig:binary}(A) and (B), respectively, show that the fast and slow SPPs mostly segregate such that the fast SPPs are highly concentrated in the patterned region and the slow SPPs are more concentrated in the non-patterned region.  In comparison, the two types of SPPs are mixed in the case where the substrate is fully uniform, as shown by Figs.~\ref{fig:binary}(C) and (D), except that the fast SPPs are more concentrated at the confining wall than the slow SPPs. 

{   The separation between the fast and slow SPPs shown in Figs.~\ref{fig:binary} (A) and (B) is counterintuitive in that the coupling between the pattern of the substrate and the motility force tend to expel the SPPs from the patterned region. Therefore, one would expect that the fast SPPs are more concentrated in the non-patterned region and that the slow SPPs are more present in the patterned region, as discussed earlier in Section III.A,  which is opposite to what is observed from Figs.~\ref{fig:binary} (A) and (B).  The fact that the patterned substrate is able to segregate the SPPs based on their motilities is very interesting and potentially very useful. However, an explanation of this phenomenon is lacking at the moment and requires further systematic simulations. This segregation could be understood from a balance of the normal stresses exerted by the SPPs at the interface between the patterned and non-patterned regions,  using for example the Irving-Kirkwood formalism~\cite{irving50}. This study is planned to be performed by the authors in the near future. Separation between SPPs may also be induced through differences in their interaction strength with the substrate and possibly the degree of their flexibility.}

\section{Summary and Conclusions}

We showed in this article that a complex collective behavior is exhibited by SPPs that are confined in a circular geometry and that interact with a circularly patterned substrate, which tends to orient the SPPs polarities with the local tangent of the pattern.  This collective behavior is characterized by SPPs vortical motion, accumulation in the outer portion of the patterned region and/or the system boundary, and SPPs exclusion from the center. This collective behavior is enhanced with increasing  SPPs driving force.  The size of the exclusion zone is determined by an interplay between, on one hand, the combined effects of the driving force and the patterned substrate, which tends to drive the SPPs outward, and, on the other hand,  motion of the SPPs  in the non-patterned region of the substrate which drives the SPPs into the patterned region. Interestingly, the vortices in the patterned and non-patterned regions, at intermediate values of the SPPs packing fraction, may have same or opposite directions.

 Another interesting feature of this system is that at intermediate packing fractions and intermediate values of the motility force, the radial profile of the packing fraction is non-monotonic, with a peak in the patterned region close to its boundary with the non-patterned region. A simulation of a binary system, composed of slow and fast SPPs (i.e., SPPs with a low and motility forces, respectively) show that they can be segregated such that the fast SPPs are mostly trapped in the patterned region, while the fast SPPs are mainly in the non-patterned region. This implies that SPPs can be segregated based on their motility.

With increasing packing fraction, 
the exclusion zone in the center disappears. High misalignment between the SPPs polarities and tangential velocities, in the center of the system, leads to an increased degree of fluctuations in their tangential velocities and reversals in the vorticity that originate from the center. Interestingly, these reversals become quasi-periodic at high packing fractions. It is worth noting that while the system exhibits vorticity reversal at both intermediate and high  packing fractions, the mechanisms leading to the two types of reversals are different. The results of the present work implies that circular patterning of the substrate can be used as a tool to guide the motion of SPPs into a collective vortical motion, and that at high packing fractions, can be used to create quasi periodic reversals in their vortical motion.

We also showed that the patterned substrate is able to segregate a binary mixture of slow and fast SPPs. We expect that SPPs can likewise be segregated based on their degrees of adhesion to the substrate. This segregation can be enhanced by further increasing the adhesion strength of the fast SPPs to the substrate.

{   We note that the present model of SPPs accounts for details often not accounted for in other models. These include elongation of the self-propelled particles, their flexibility, and enclosed area of the SPPs. It would of course be very desirable to determine the effects of each of these ingredients on the details of the results. 
There is of course a close connection between the SPP dynamics described here with that of swimming bacteria. However, it is important to note that the
estimated value of the Reynolds number based on the parameters used in this study (Eq.~(15)) is about 1, which is much larger than that of swimming bacteria. Using the present approach to investigate the collective motion of cells such as bacteria requires a much smaller Reynolds number which can be achieved by increasing the value of the drag coefficient $\Gamma$ in our model. We plan to investigate the effects of these parameters on the observed phenomena in the present study in the near future.}

\section{Acknowledgements}
All simulations were performed on computers of the High Performance Computing Facility of the University of Memphis. This work was funded by the University of Memphis.

\end{document}